%% file: main.tex
\PassOptionsToPackage{table,xcdraw}{xcolor}
\documentclass[sigconf]{acmart}

\usepackage{booktabs}
\usepackage{tabularx}
\usepackage{multirow}
\usepackage{colortbl}
\usepackage{enumitem} 

\renewcommand\footnotetextcopyrightpermission[1]{} 

\acmBooktitle{}
\acmYear{}
\acmISBN{}
\acmDOI{}

\input{_global_settings}

\begin{document}

\title{A Large-Scale Evolvable Dataset for Model Context Protocol Ecosystem and Security Analysis}

\author{Zhiwei Lin\hspace{0.3em}\ Bonan Ruan\hspace{0.3em}\ Jiahao Liu\hspace{0.3em}\ Weibo Zhao}
\affiliation{%
  \institution{\vspace{0.1cm}\ National University of Singapore}
  \city{}
  \country{}
}
\vspace{0.1cm}
\email{{zhiweil, r-bonan, jiahao99, weibo}@comp.nus.edu.sg}

\renewcommand{\shortauthors}{Zhiwei Lin, Bonan Ruan, Jiahao Liu, Weibo Zhao}

\input{sections/abstract}

\maketitle

\input{sections/introduction}

\input{sections/dataset_construction}
\input{sections/dataset_description}
\input{sections/application}

\input{sections/conclusion}
%

\bibliographystyle{ACM-Reference-Format}
\bibliography{ref}


\end{document}

%% file: _global_settings.tex
\usepackage{dingbat}
\usepackage{listings}
\usepackage{color}
\usepackage{multirow}
\usepackage{colortbl}
\usepackage{enumitem}

\usepackage[linesnumbered,ruled,boxed,lined]{algorithm2e}
\SetAlgoVlined 
\SetAlgoLined 
\SetKwInput{KwInput}{Input}  
\SetKwInput{KwOutput}{Output}  
\SetKwProg{Proc}{Procedure}{}{}

\DontPrintSemicolon

\usepackage{algpseudocode}
\usepackage{url}
\usepackage{float}
\usepackage{hyperref}
\usepackage{caption}
\usepackage{adjustbox}

\algtext*{EndProcedure}

\definecolor{mypurple}{HTML}{67379A}
\definecolor{myyellow}{HTML}{FAD978}
\definecolor{myblue}{HTML}{93AAD8}
\definecolor{myred}{HTML}{EB3323}
\definecolor{mygreen}{HTML}{4EAD5B}
\definecolor{greenbg}{rgb}{0.9, 1, 0.9}

\lstdefinestyle{mystyle}{
    backgroundcolor=\color{white},   
    basicstyle=\ttfamily\footnotesize,
    breakatwhitespace=false,         
    breaklines=true,                 
    captionpos=b,                    
    keepspaces=true,                 
    showspaces=false,                
    showstringspaces=false,
    showtabs=false,                  
    tabsize=2,
    frame=single, 
    framerule=0pt,
    xleftmargin=5mm,
    xrightmargin=3mm,
    framesep=5mm, 
    rulecolor=\color{black},
    postbreak=\mbox{\textcolor{red}{$\hookrightarrow$}\space},
}
\lstdefinelanguage{json}{
    basicstyle=\ttfamily\footnotesize,
    showstringspaces=false,
    breaklines=true
}

\usepackage{amsthm}
\theoremstyle{definition}
 
\theoremstyle{plain}

\newcommand{\lgtm}{\textsc{MCPCorpus}\xspace}

\usepackage{pifont}

\makeatletter
\newcommand{\removelatexerror}{\let\@latex@error\@gobble}
\makeatother

%% file: sections/abstract.tex
\begin{abstract}

The Model Context Protocol (MCP) has recently emerged as a standardized interface for connecting language models with external tools and data.
As the ecosystem rapidly expands, the lack of a structured, comprehensive view of existing MCP artifacts presents challenges for research.
To bridge this gap, we introduce \lgtm, a large-scale dataset containing around 14K MCP servers and 300 MCP clients. Each artifact is annotated with 20+ normalized attributes capturing its identity, interface configuration, GitHub activity, and metadata.
\lgtm provides a reproducible snapshot of the real-world MCP ecosystem, enabling studies of adoption trends, ecosystem health, and implementation diversity.
To keep pace with the rapid evolution of the MCP ecosystem, we provide utility tools for automated data synchronization, normalization, and inspection.
Furthermore, to support efficient exploration and exploitation, we release a lightweight web-based search interface.
\lgtm is publicly available at: \href{https://github.com/Snakinya/MCPCorpus}{https://github.com/Snakinya/MCPCorpus}.

\end{abstract}

%% file: sections/introduction.tex
\section{Introduction}

Given their strong capabilities in language understanding, reasoning, and decision-making, large language models (LLMs) have been increasingly integrated with diverse tools and services to collaboratively perform a wide range of tasks~\cite{shen2024llm,naveed2023comprehensive}.

However, due to the heterogeneity of tool interfaces and the lack of standardized integration protocols, swiftly integrating them remains challenging, which significantly hinders the development of LLM-based applications~\cite{nahar2024beyond}.
To address this challenge, Model Context Protocol (MCP) has been introduced as a lightweight, standardized interface that connects LLMs with diverse tools~\cite{mcp2025spec}.
Specifically, MCP defines tools as callable endpoints with structured JSON schemas, allowing developers to expose services in a consistent and interoperable manner across different programming environments and platforms.
Additionally, MCP employs a modular client-server architecture, where the MCP server registers tool capabilities and executes them upon request, while the MCP client discovers these capabilities and invokes the desired tools through a standardized protocol~\cite{singh2025survey}.

Although standardization has driven the rapid growth of MCP, fostering an active and expansive ecosystem, it has also exposed a range of issues, including inconsistent implementations, malicious server behaviors, and weak protocol compliance~\cite{maliciousmcp,hou2025model}.
This phenomenon has shifted the focus of research from merely exploring new applications of MCP to examining its reliability, security, and ecosystem dynamics~\cite{fang2025we,song2025protocolunveilingattackvectors}.
However, we identify several challenges in existing studies:
(a) \textit{Limited-source investigation}. Most works focus on MCP artifact information from a single source,  ignoring the rich metadata available across different platforms such as code hosting sites (e.g., GitHub), community hubs (e.g., \texttt{MCP.so}~\cite{mcpServers}), and package managers.
(b) \textit{Unscalable analysis}. These studies often rely on hand-picked examples, lacking the scalability needed for a comprehensive assessment of the MCP ecosystem’s current state.
The underlying reason is the absence of a large-scale, well-archived MCP dataset that consolidates MCP servers from diverse sources and unifies them for ease of use.

To bridge this gap and better support future research on the MCP ecosystem, we present \lgtm, a large-scale dataset accompanied by utility scripts for maintaining and continuously updating the dataset.
\lgtm provides a comprehensive view of real-world MCP artifacts, currently including around 14K MCP servers and 300 MCP clients.
For a clear overview and ease of analysis, each MCP artifact in \lgtm is annotated with over 20 metadata attributes, including \textit{type}, \textit{tool list}, \textit{programming language}, \textit{license}, \textit{author name}, \textit{server configuration}, and \textit{GitHub activity indicators}.
Take \textit{tool list} as an example: it allows users to quickly understand the capabilities provided by an MCP server and how to interact with it, helping them efficiently determine whether the server is relevant for a given investigation.
It is important to note that these attributes are extracted through static inspection of public repositories, without requiring runtime access or live service interaction.

With this dataset, we envision a range of potential applications, including (1) interoperability benchmarking of tool-augmented LLM agents, (2) security auditing of MCP implementations via linked source code analysis, and (3) schema conformance checking across diverse MCP artifacts. For example, researchers can locate implementations through metadata and inspect code to identify issues like missing authentication or unsafe endpoint exposure.

In summary, we make the following contributions:
\begin{itemize}[leftmargin=12pt]
\item We curate a large-scale dataset named \lgtm, covering around 14K MCP servers and 300 clients from multiple sources. These artifacts span diverse domains and programming languages, and are annotated with 20+ normalized attributes to support ecosystem-level analysis.

\item We develop a set of utility tools to support data synchronization, normalization, and inspection, as well as a public-facing web interface to explore the dataset.
\end{itemize}

%% file: sections/dataset_construction.tex
\section{Dataset Construction}
\label{sec:construction}

To ensure that \lgtm can comprehensively profile MCP-related artifacts and support various research and engineering tasks, we conduct a systematic investigation of the current MCP ecosystem and integrate information from multiple online sources.
In particular, we focus on metadata-rich, actively maintained platforms that cover a wide range of MCP servers and clients.

In the subsequent sections, we first introduce the data sources selected for \lgtm and then elaborate on the data collection process, including our crawling strategy, metadata normalization, and attribute synthesis.

\subsection{Data Sources}
To construct a large-scale and high-quality dataset of MCP implementations, we combine two complementary sources: MCP.so, which hosts the largest and most information-rich collection of MCP tools, and GitHub, which provides additional implementation and maintenance metadata.

\textbf{MCP.so} is currently the most extensive registry dedicated to MCP servers and clients.
As of 3 June 2025, \texttt{MCP.so} hosts over 14,000 MCP servers and 300 client implementations, and continues to update its database daily. Compared to other MCP hosting platforms, \texttt{MCP.so} offers the most comprehensive artifact-level information, including textual descriptions, domain tags, author identities, publication dates, and deployment statuses. To enable reliable analysis of the MCP ecosystem, we developed a unified metadata schema and systematically extracted and normalized over ten key attributes from \texttt{MCP.so}'s unstructured content.

\textbf{GitHub} complements \texttt{MCP.so} by providing rich technical metadata for each associated repository.
We extract statistics such as star count, forks, open issues, and contributors, as well as signals of activity and maintenance (\textit{e.g.,} last commit time and license type).
We also detect technical traits like programming language composition, and the presence of Dockerfile and README files, which help enrich the static MCP registry entries with development context.

\subsection{Data Collection}
\label{sec:data_collection}

\begin{figure}[t]
  \centering
  \includegraphics[width=0.8\linewidth]{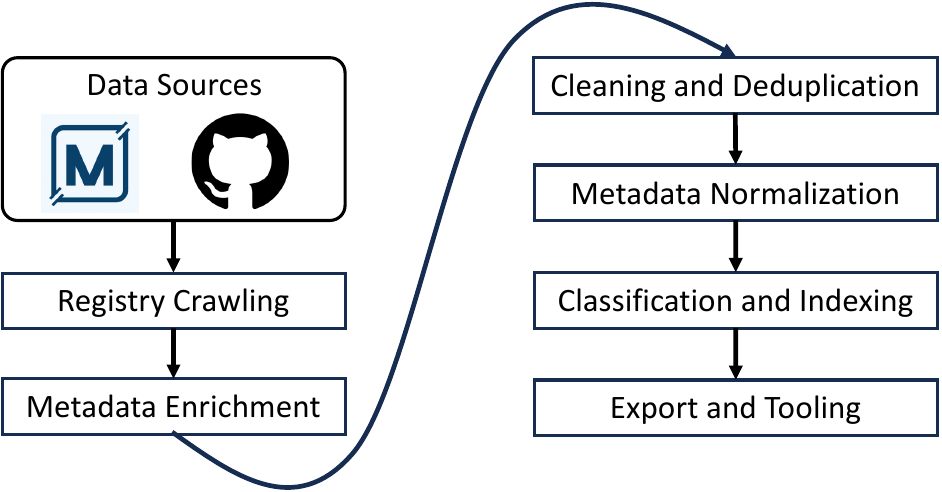}
  \caption{Approach Overview}
  \label{fig:overview}
\end{figure}

The construction of \lgtm involves a five-stage pipeline to extract, enrich, normalize, and classify MCP artifacts from both centralized registries and decentralized code repositories.
As shown in \autoref{fig:overview}, this process is designed to ensure the dataset is comprehensive, clean, and useful for both research and engineering tasks.
The key stages are detailed as follows.

\textbf{Registry Crawling.}  
The collection process begins with crawling \texttt{MCP.so}.
Our crawler systematically traverses its paginated listings and artifact detail pages to extract structured metadata, including artifact name, description, domain tags, category, tool schema, author information, deployment interface (\textit{e.g.,} REST APIs), creation/update timestamps, and GitHub URLs.

A total of around 13.9K MCP servers and 300 clients were extracted at the time of collection.
For example, one of the MCP server projects, \textit{edgeone-pages-mcp}, exposes a tool named \texttt{deploy-html} and is tagged under the \texttt{cloud-platforms} category, with a GitHub repository maintained by TencentEdgeOne.

\textbf{GitHub Metadata Enrichment.}  
For each MCP artifact with a valid GitHub link, we query the GitHub REST API to collect its repository-level attributes. These include popularity metrics (\textit{e.g.,} \textit{stars}, \textit{forks}), maintenance signals (\textit{e.g.,} \textit{last commit time}, \textit{contributor count}), \textit{license type}, and the composition of programming languages (\textit{e.g.,} TypeScript, JavaScript). We also analyze the file structure to detect critical assets like \textit{README.md}, \textit{Dockerfile}, and other deployment-related files. For the aforementioned example, \textit{edgeone-pages-mcp}, this phase confirms 125 stars, 16 forks, an MIT license, and recent activity (last commit in May 2025).

\textbf{Cleaning and Deduplication.}  
To ensure data quality, we remove artifacts with invalid GitHub links, unreachable repositories, or insufficient metadata—defined as missing critical fields such as repository URLs.
Duplicate detection is based on GitHub repository canonicalized URLs. If multiple entries point to the same repository (e.g., forks, mirrors, or renamed variants), we retain the version with the highest activity level (e.g., stars, recent commits) or the earliest creation timestamp as a proxy for originality.
This process significantly reduces noise and ensures dataset consistency.

\textbf{Metadata Normalization.}  
Each artifact’s metadata is parsed and normalized into a unified schema with 20+ attributes.
These attributes include basic descriptors (\textit{e.g.,} category, interface type), technical metadata (\textit{e.g.,} programming language, license), and derived indicators (\textit{e.g.,} maintenance status, tool count). Fields originally encoded as JSON within the registry (\textit{e.g.,} tool definitions, custom metadata blocks) are parsed and flattened into structured representations, enabling efficient downstream querying, filtering, and analysis.
As not all attributes are applicable to every artifact type (\textit{e.g.,} clients do not define tools), missing or inapplicable fields are filled with \texttt{null}.

\textbf{Classification and Indexing.}  
To facilitate structured exploration, each artifact is categorized along three core dimensions that are currently implemented: (i) artifact type (server or client), (ii) application category (e.g., cloud, AI, data), and (iii) programming language.
These basic labels enable queries such as “Python-based AI servers” or “Go-written data clients.”
Additional dimensions such as interface type, officiality, and maintenance status are part of our design and will be integrated in future versions to support more fine-grained analysis.

\textbf{Export and Tooling.}  
The final dataset is stored in newline-delimited JSON (JSONL) format. Each record contains merged registry and GitHub metadata along with classification tags. We also provide auxiliary tools for data synchronization, statistical analysis, and web-based exploration. These utilities enable reproducible updates and promote ease of integration into research pipelines.

%% file: sections/dataset_description.tex
\section{Dataset Description}
\label{sec:description}

\subsection{Basic Composition}

\definecolor{basiccolor}{HTML}{E8EAF6}
\definecolor{interfacecolor}{HTML}{E0F2F1}
\definecolor{signalscolor}{HTML}{FFF9C4}
\definecolor{metadatacolor}{HTML}{FCE4EC}

\begin{table}[t]
\caption{Refined field schema of the \lgtm dataset.}
\label{tab:fields_final_integrated}
\begin{tabularx}{\columnwidth}{@{} >{\bfseries\ttfamily}l X @{}}
\toprule

\textbf{Field} & \textbf{Description} \\
\midrule

\rowcolor{basiccolor}
id & Unique identifier of the record \\
\rowcolor{basiccolor}
name & Short name of the MCP artifact \\
\rowcolor{basiccolor}
title & Display title of the project \\
\rowcolor{basiccolor}
description & Human-readable summary \\
\rowcolor{basiccolor}
author\_name & GitHub username or organization \\
\rowcolor{basiccolor}
url & Repository URL \\
\rowcolor{basiccolor}
category & Domain category (\textit{e.g.,} AI, tools) \\
\rowcolor{basiccolor}
tags & Comma-separated topic labels \\
\rowcolor{basiccolor}
type & Whether the artifact is a server or client \\

\rowcolor{interfacecolor}
tools & MCP tools or functions exposed (JSON) \\
\rowcolor{interfacecolor}
sse\_url & Callable endpoint if applicable \\
\rowcolor{interfacecolor}
server\_command & Execution command or entry point \\
\rowcolor{interfacecolor}
server\_config & Tool config and environment variables \\

\rowcolor{signalscolor}
stargazers\_count & GitHub stars \\
\rowcolor{signalscolor}
forks\_count & GitHub forks \\
\rowcolor{signalscolor}
open\_issues\_count & Number of open issues \\
\rowcolor{signalscolor}
contributors\_count & Number of unique contributors \\
\rowcolor{signalscolor}
last\_commit & Timestamp of the most recent commit \\

\rowcolor{metadatacolor}
full\_name & Full GitHub repo name (\textit{e.g.,} user/project) \\
\rowcolor{metadatacolor}
language & Primary implementation language \\
\rowcolor{metadatacolor}
languages & Byte-level language breakdown \\
\rowcolor{metadatacolor}
license & License type (\textit{e.g.,} MIT) \\
\rowcolor{metadatacolor}
archived & Whether the repository is archived \\
\rowcolor{metadatacolor}
has\_docker & Presence of a Dockerfile \\
\rowcolor{metadatacolor}
has\_readme & Presence of README.md \\
\rowcolor{metadatacolor}
has\_requirements & Presence of dependency declarations \\
\bottomrule
\end{tabularx}

\vspace{4pt}
\begin{flushleft}
\small
\textcolor{basiccolor}{\rule{6pt}{6pt}}~Basic Info | 
\textcolor{interfacecolor}{\rule{6pt}{6pt}}~Interface \& Config | 
GitHub \{ \textcolor{signalscolor}{\rule{6pt}{6pt}}~Signals, 
\textcolor{metadatacolor}{\rule{6pt}{6pt}}~Metadata \}
\end{flushleft}
\end{table}

The \lgtm dataset consists of around 14K MCP artifacts, including 13,875 servers and 300 clients. Each artifact is represented as a JSON object with up to 26 structured attributes, combining metadata from \texttt{MCP.so} and enriched signals from GitHub repositories. These attributes describe the artifact’s identity, deployment interface, functionality, language stack, license, and development activity. All entries conform to a unified schema, supporting consistent parsing and downstream processing.

\subsection{Attribute Categorization}

To support downstream analysis and improve the interpretability of the dataset, we group the refined fields into four functional categories based on their semantic roles and data sources.
This categorization helps researchers and developers better understand the structure of each artifact and selectively utilize relevant attributes for specific tasks such as filtering, compatibility checking, and repository quality assessment.
\autoref{tab:fields_final_integrated} provides an overview of all retained fields and their corresponding descriptions.

\begin{itemize}[leftmargin=12pt]
    \item \textbf{Basic Information}: Fields such as \textit{id}, \textit{name}, \textit{title}, \textit{description}, \textit{author\_name}, \textit{url}, \textit{category}, \textit{tags}, and \textit{type} provide fundamental metadata that describes the identity, content, and domain classification of the MCP artifact.
    
    \item \textbf{Interface and Configuration}: Fields such as \textit{tools}, \textit{sse\_url}, \textit{server\allowbreak\_command}, and \textit{server\_config} capture how an MCP artifact exposes its functionality, including runtime interface details and tool execution configuration. These fields are critical for tool compatibility analysis and endpoint validation.
    
    \item \textbf{GitHub Signals}: Fields such as \textit{stargazers\_count}, \textit{forks\allowbreak\_count}, \textit{open\_issues\_count}, \textit{contributors\_count}, and \textit{last\allowbreak\_commit} represent quantitative signals extracted from GitHub that reflect project popularity, development activity, and community engagement. They are useful for analyzing development characteristics and identifying active or well-supported artifacts.
    
    \item \textbf{GitHub Metadata}: Fields such as \textit{full\_name}, \textit{language}, \textit{languages}, \textit{license}, \textit{archived}, \textit{has\_docker} and \textit{has\_readme} capture the technical structure, licensing, and documentation quality of each artifact. These attributes are valuable for security auditing, codebase profiling, and integration feasibility studies.
\end{itemize}

\subsection{Data Sample}

To illustrate the structure and metadata richness of individual entries in \lgtm, we present a representative MCP server named \textit{playwright-mcp}, developed and maintained by Microsoft~\cite{githubGitHubMicrosoftplaywrightmcp}.
This artifact belongs to the \textit{browser-automation} category and is implemented in TypeScript. It exposes a command-line interface for headless browser control via Playwright, with metadata drawn from both the \texttt{MCP.so} registry and the corresponding GitHub repository.
A simplified version of the entry is shown in \autoref{lst:mcp_entry}.
This sample contains all 26 structured fields defined in \lgtm's schema, including both registry metadata (\textit{e.g.,} category, tags, server command) and GitHub-derived indicators (\textit{e.g.,} stars, forks, contributors).
Even if some values are absent (\textit{e.g.,} \texttt{sse\_url}), the corresponding keys are retained to preserve consistency across records.

\begin{figure}[t]
\begin{lstlisting}[
    language=json,
    basicstyle=\footnotesize\ttfamily,
    frame=single,
    breaklines=true,
    showstringspaces=false,
    belowskip=-0.3\baselineskip
]
{ "id": 4493, "name": "playwright-mcp", "title": "Playwright Mcp",
  "url": "https://github.com/microsoft/playwright-mcp",
  "description": "Playwright MCP server", "author_name": "microsoft",
  "tags": "mcp,playwright", "category": "browser-automation",
  "type": "server", "tools": "", "sse_url": null,
  "server_command": "docker exec -i mcp-node bash -c \"npx @playwright/mcp@latest --headless\"",
  "server_config": {"mcpServers": {"playwright": {"command": "npx",
  "args": ["@playwright/mcp@latest","--headless"]}}},
  "github": {"full_name": "microsoft/playwright-mcp", "forks_count": 727,
  "stargazers_count": 11162, "open_issues_count": 22, "contributors_count": 24,
  "language": "TypeScript", "languages": {"TypeScript": 188386, 
  "JavaScript": 13482, "Dockerfile": 2210}, "license": "Apache License 2.0",
  "archived": false, "has_docker": true, "has_readme": true,
  "has_requirements": false, "last_commit": "2025-06-03T18:10:47Z"}}
\end{lstlisting}
\caption{Full MCP server entry covering all schema fields.}
\label{lst:mcp_entry}
\end{figure}

\subsection{Insights and Characteristics}

To further understand the structure and dynamics of the MCP ecosystem, we analyze two critical characteristics from the \lgtm dataset: GitHub stargazer count and primary implementation language. These two features not only reflect project popularity and technical preference but also exhibit diverse distributions that support meaningful ecosystem-level insights.

\begin{figure}[t]
    \centering
    \includegraphics[width=\linewidth]{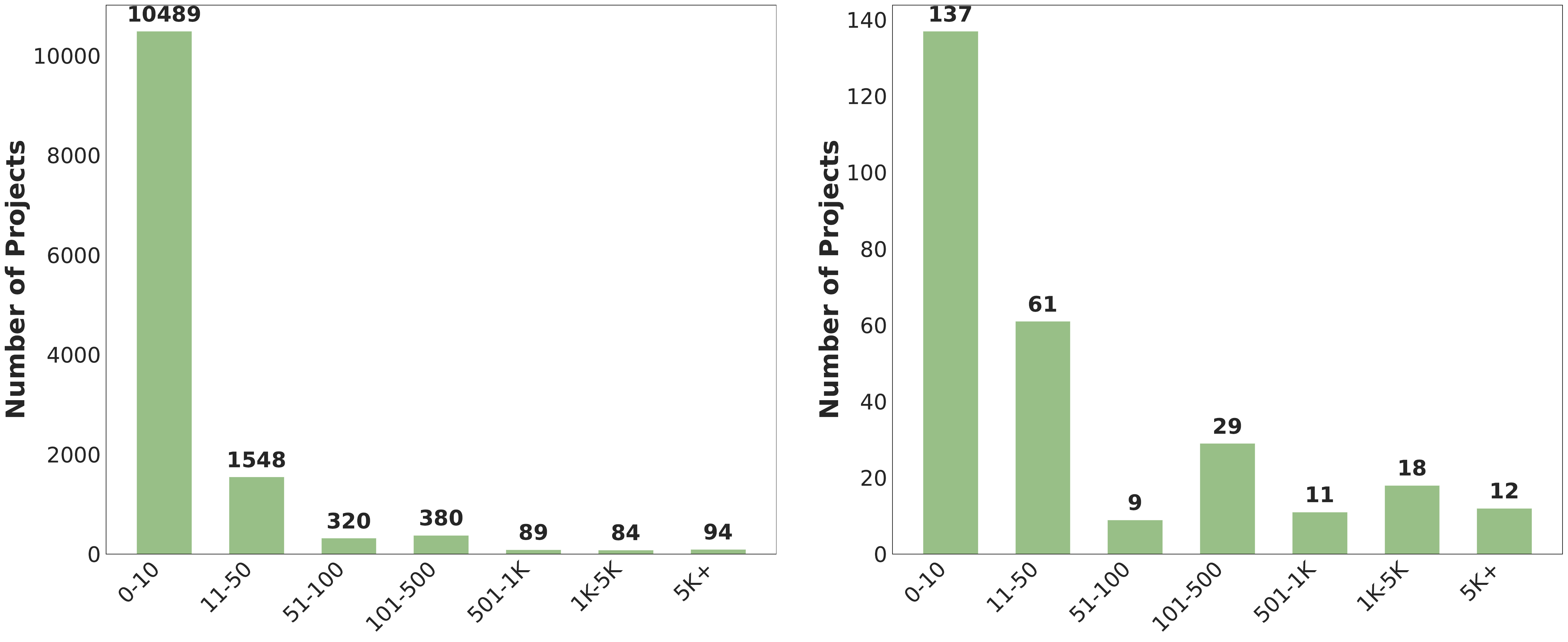}
    \caption{Distribution of GitHub stars for all MCP servers (left) and MCP clients (right).}
    \label{fig:stars}
\end{figure}

\textbf{Popularity Distribution.} \autoref{fig:stars} illustrates the distribution of GitHub stars across MCP artifacts. We observe a highly skewed long-tail pattern: the majority of projects (over 10,000) have fewer than 10 stars, while only a small fraction exceed 500 or 1,000 stars. This suggests that although the ecosystem is rapidly growing, most MCP projects remain in early stages of development or adoption. Notably, some highly starred projects surpass 5,000 stars, indicating early consolidation of a few popular tools.

\begin{figure}[t]
    \centering
    \includegraphics[width=\linewidth]{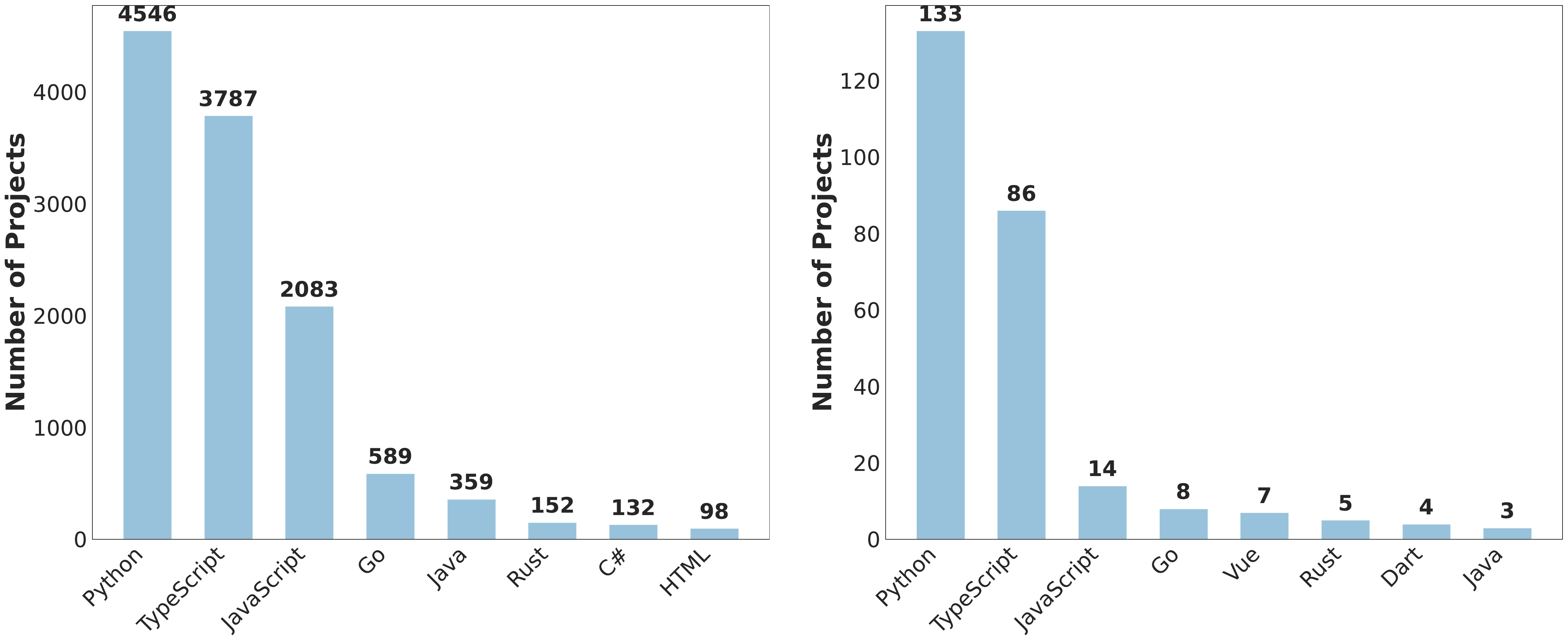}
    \caption{Primary programming language distribution for all MCP servers (left) and MCP clients (right).}
    \label{fig:langs}
\end{figure}

\textbf{Language Ecosystem.} \autoref{fig:langs} shows the distribution of primary programming languages among MCP artifacts.
Specifically, Python, TypeScript, and JavaScript are the most commonly used languages, accounting for the majority of implementations.
Languages such as Go, Rust, Java, and C\# also appear but with lower frequency. Interestingly, the distribution is consistent even when filtered to the top-starred projects, suggesting that these languages are not only popular but also dominant among well-maintained, widely adopted MCP artifacts.

These observations confirm that \lgtm captures both the breadth and structural concentration of the current MCP ecosystem, enabling downstream studies on language-specific adoption, sustainability, and integration practices.

%% file: sections/application.tex
\section{Application Scenarios}
\label{sec:application}

\subsection{Ecosystem Analysis}
\lgtm enables a systematic analysis of the MCP ecosystem by aggregating structured metadata across a diverse set of MCP server implementations.
Through standardized fields such as language, license, last\_commit, and contributors\_count, researchers can investigate trends in programming language adoption, project vitality, and open-source collaboration patterns. Additionally, fields like created\_at, updated\_at, and stargazers\_count provide temporal insights into growth dynamics and community interest. This allows longitudinal tracking of the ecosystem's expansion and supports ecosystem health evaluation, including identifying under-maintained or overly centralized projects. Overall, \lgtm functions as a longitudinal data source to support quantitative ecosystem modeling for LLM-integrated infrastructure.

\subsection{Security and Code Quality Assessment}
Beyond ecosystem analysis, \lgtm facilitates research on the security posture and code quality of MCP servers and clients by exposing their runtime interface details and repository characteristics.
Fields such as server\_command, server\_config, allow\_call, and tools reveal the external interface surface, which can be leveraged to evaluate attack exposure and command execution behavior. Metadata like has\_docker, has\_readme, and forks\_count reflects engineering completeness and adoption footprint, serving as auxiliary signals for maintainability and documentation hygiene.
Combined with language and license information, \lgtm supports automated vulnerability analysis, dependency auditing, and cross-project policy enforcement.
The dataset can serve as a benchmark corpus for empirical studies in secure-by-design MCP development and for training tools that assess AI-agent interoperability risks.

%% file: sections/conclusion.tex
\section{Conclusion}
\label{sec:conclusion}

In this paper, we introduce \lgtm, a large-scale and evolvable dataset capturing the structure and dynamics of the MCP ecosystem.
By integrating metadata from diverse sources, \lgtm offers a unified view of MCP artifacts, supporting analyses of interoperability, security, and ecosystem health.
The dataset, along with utility tools, is publicly released to foster reproducible research and inform the development of robust, tool-augmented LLM systems.